\documentclass[aps,prl,nobibnotes,twocolumn,superscriptaddress,bibliography]{revtex4-2}

\usepackage{chemformula} 
\usepackage[T1]{fontenc} 

\usepackage{amsfonts}
\usepackage{mathrsfs}
\usepackage{amsmath}
\usepackage{color}
\usepackage{graphicx}
\graphicspath{ {./pictures/version_6/} }
\usepackage{bm}
\usepackage{amssymb}
\usepackage{xspace}
\usepackage{epstopdf}
\usepackage{dcolumn}
\usepackage{longtable}
\usepackage{multirow}
\usepackage{float}
\usepackage{comment}
\usepackage{lipsum}
\usepackage{tabularx}
\usepackage{booktabs}
\def\ie{{\it i.e.},\ }

\usepackage{physics}

\usepackage[colorlinks=true, letterpaper=true, pdfstartview=FitV, linkcolor=blue, citecolor=blue, urlcolor=blue]{hyperref}

\newcommand{\Rom}[1]{\uppercase\expandafter{\romannumeral#1}}

\makeatother

\begin{document}
\title
{Two-dimensional Topological Ferroelectric Metal with Giant Shift Current}

\author{Liu Yang}
\affiliation
{Department of Physics, Hubei Engineering Research Center of Weak Magnetic-field Detection, China Three Gorges University, Yichang, 443002, China}

\author{Lei Li}
\email[Liu Yang and Lei Li contributed equally to this work\\]{lilei1993@bit.edu.cn}
\affiliation{Centre for Quantum Physics, Key Laboratory of Advanced Optoelectronic Quantum Architecture and Measurement (MOE), School of Physics, Beijing Institute of Technology, Beijing, 100081, China}
\affiliation{Beijing Key Lab of Nanophotonics \& Ultrafine Optoelectronic Systems, School of Physics, Beijing Institute of Technology, Beijing, 100081, China}

\author{Zhi-Ming Yu}
\affiliation{Centre for Quantum Physics, Key Laboratory of Advanced Optoelectronic Quantum Architecture and Measurement (MOE), School of Physics, Beijing Institute of Technology, Beijing, 100081, China}
\affiliation{Beijing Key Lab of Nanophotonics \& Ultrafine Optoelectronic Systems, School of Physics, Beijing Institute of Technology, Beijing, 100081, China}
\affiliation{International Center for Quantum Materials, Beijing Institute of Technology, Zhuhai, 519000, China}

\author{Menghao Wu}
\email{wmh1987@hust.edu.cn}
\affiliation
{School of Physics, Huazhong University of Science and Technology, Wuhan, 430074, China}

\author{Yugui Yao}
\email{ygyao@bit.edu.cn}
\affiliation{Centre for Quantum Physics, Key Laboratory of Advanced Optoelectronic Quantum Architecture and Measurement (MOE), School of Physics, Beijing Institute of Technology, Beijing, 100081, China}
\affiliation{Beijing Key Lab of Nanophotonics \& Ultrafine Optoelectronic Systems, School of Physics, Beijing Institute of Technology, Beijing, 100081, China}
\affiliation{International Center for Quantum Materials, Beijing Institute of Technology, Zhuhai, 519000, China}

\begin{abstract}
The pursuit for "ferroelectric metal" which combines seemingly incompatible spontaneous electric polarization and metallicity, has been assiduously ongoing but remains elusive. Unlike traditional ferroelectrics with a wide band gap, ferroelectric (FE) metals can naturally incorporate nontrivial band topology near the Fermi level, endowing them with additional exotic properties. Here, we show first-principles evidence that the metallic \ch{PtBi2} monolayer is an intrinsic two-dimensional (2D) topological FE metal, characterized by out-of-plane polarization and a moderate switching barrier. Moreover, it exhibits a topologically nontrivial electronic structure with $\mathbb Z_2$ invariant equal to 1, leading to a significant FE bulk photovoltaic effect. A slight strain can further enhance this effect to a remarkable level, which far surpass that of previously reported 2D/3D FE materials. Our work provides an important step towards realizing intrinsic monolayer topological FE metals and paves a promising way for future nonlinear optical devices.

\end{abstract}

\maketitle

\begin{figure*}
	\begin{centering}
		\includegraphics[width=\textwidth]{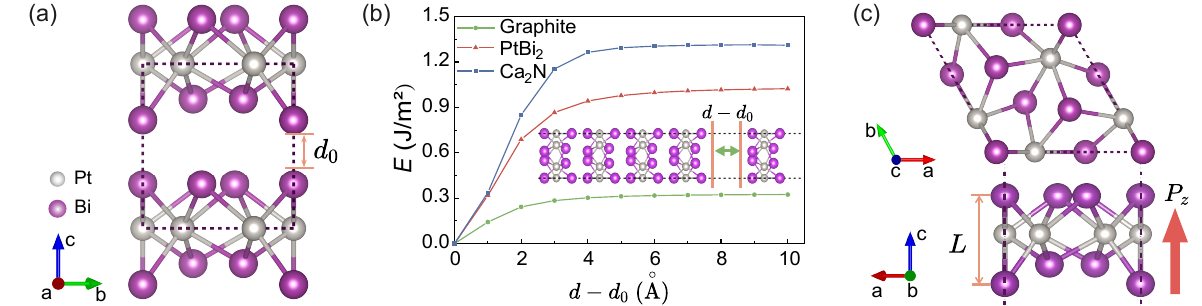}
		\par\end{centering}
	\caption{
		(a) Crystal structure of bulk t-\ch{PtBi2}. Pt and Bi atoms are labeled with gray and purple colors, respectively, and $d_0$ denotes the equilibrium interlayer spacing in the bulk state.
		(b)  Calculated exfoliation energy versus separation distance in comparison with graphite and \ch{Ca2N}.
		(c) Top and side views of monolayer \ch{PtBi2}, with a thickness of $L = 4.06~\text{\AA}$.
	}
	\label{fig:fig1}
\end{figure*}

\textcolor{blue}{\textit{Introduction.}}---
Ferroelectric materials, characterized by their spontaneous polarization that can be switched by an external electric field, hold tremendous potential for a wide range of applications, including non-volatile memories, transducers, nonlinear optical devices, and other innovative electronic devices  \cite{RevModPhys.77.1083,scienceScott,adma.202005098,AdvEnergyMat}.
Traditionally, ferroelectricity and metallicity have been considered incompatible, as the itinerant electrons in metals can effectively screen electric fields and dipoles.
In 1965, Anderson and Blount  \cite{Anderson} suggested a scenario where a metallic state could undergo a ferroelectric-like structural transition, marked by the appearance of a polar axis and the loss of an inversion center.
This theoretical prediction was experimentally validated in bulk \ch{LiOsO3}   \cite{shi2013ferroelectric}, which exhibited a temperature-driven transition from a centrosymmetric ($R\bar{3}c$) to a non-centrosymmetric ($R3c$) structure. 
However, the two ferroelectric-like states cannot yet be switched by an electric field, and these materials are referred as "polar metals"    \cite{kim2016polar,C5TC03856A,Zhou_2020,PhysRevLett.115.087202}.

In two-dimensional (2D) materials, however, conduction electrons are confined within the plane, allowing an external electric field to penetrate the material and reverse its out-of-plane polarization. Base on this idea, several 2D ferroelectric (FE) metal following different design principles have been progressively proposed in the literature  \cite{PRBXiang,PRLXiang,GangSu,MaterHorizLiuShi,ai2022FE,PhysRevB.108.104109,PhysRevB.109.035421,yu2024two}, but the corresponding experimental validation remains elusive.
The $1\text{T}^\prime$-\ch{WTe2} multilayer is the first FE metal verified experimentally \cite{fei2018ferroelectric}, with its polarization directly detected and quantified using graphene as an electric-field sensor.
To date, the identified 2D FE metals are still rare, and to the best of our knowledge, no intrinsic monolayer FE metal has been experimentally confirmed.

Owing to the broken inversion symmetry, ferroelectrics with spontaneous polarization can be promising candidates for the bulk photovoltaic effect (BPVE), a second-order nonlinear optical (NLO) process that enables the direct conversion of light into electricity \cite{ChemPhyRev,AdvOptMaterials}.
Unlike conventional photovoltaic cells, which rely on p-n junctions, the BPVE can generate an open-circuit photovoltage higher than the band gap \cite{spanier2016power}, offering a potential pathway to surpass the Shockley-Queisser limit \cite{Shockley,RUHLE2016139}.
The shift current mechanism \cite{Sipe2000,Sipe2010} is believed to be the primary intrinsic contributor to the BPVE and has been extensively observed experimentally in ferroelectrics \cite{ScienceAdvances,CuInP2S6, 3R_MoS2, AdvEnergyMat}, but how to achieve a large shift current is still an open question \cite{tan2016shift,cook2017design}.

Nontrivial band topology may play an important role in addressing this issue. 
It has been pointed out that band inversion can effectively enhance optical responses \cite{lijuJPCL,PhysRevLett.119.266804} and lead to a significant shift current \cite{PhysRevLett.116.237402,PhysRevLett.130.256902}.
Notably, the shift current of Weyl points \cite{PhysRevXNagaosa} and nodal lines \cite{PhysRevB.107.035114} is divergent at low frequencies.
However, traditional 3D FE materials, such as transition metal perovskites, are typical insulators \cite{lines2001principles}. 
The presence of a wide band gap prevents the compatibility of these materials with band topology near the Fermi level, a challenge often referred to as the “band gap dilemma” in some literature \cite{huang2022demand,liu2016strain}.
Obviously, 2D FE metals naturally circumvent this difficulty, and the coexistence of topological orders or emergent fermions \cite{YUScienceBulletin,PhysRevB.105.085117,PhysRevB.105.104426}, 
and ferroelectricity in 2D topological FE metals (semimetals) may render unique advantages in the fields of photovoltaics and photodetection.

In this work, based on the first-principles calculations, we propose that monolayer \ch{PtBi2}, characterized by a sizable spontaneous polarization and moderate switching barrier, can be an ideal candidate for intrinsic monolayer topological FE metal.
Two FE ground states with opposite polarization can be achieved through the softening of a zone-center phonon mode labeled $B_{1u}$ in a centrosymmetric non-polar state.
The electronic structure of the monolayer \ch{PtBi2} exhibits strong metallicity and possesses a topological invariant $\mathbb{Z}_2 = 1$.
Moreover, this nontrivial band structure endows it with a giant FE bulk photovoltaic effect, which can be further enhanced by a slight strain with an unprecedented shift current.

\textcolor{blue}{\textit{Topological Ferroelectric metal.}}---
Trigonal \ch{PtBi2} (t-\ch{PtBi2}), with space group $P31m$ (no. 157), is a layered van der Waals materials that can be easily synthesized in experiments \cite{zaac.201400331,PhysRevB.94.165119,PhysRevB.94.235140,PhysRevB.97.035133,nie2020robust}.
A giant linear magnetoresistance \cite{APL1.4954272} has been observed in bulk t-\ch{PtBi2} which is a semimetal with complex Fermi surface composed of multiple electron and hole pockets \cite{PhysRevB.94.165119}.
Its electronic structure also attracts considerable interest and a range of intriguing properties have been revealed, including giant anisotropic Rashba-like spin splitting \cite{feng2019rashba}, triply degenerate point fermions \cite{gao2018possible} and type-I Weyl nodes \cite{veyrat2023berezinskii} close to the Fermi level. 
More interestingly, the bulk \cite{PRM.4.124202,PhysRevB.103.014507,10.1063/10.0014014}, surface \cite{schimmel2023high} and thin flakes \cite{veyrat2023berezinskii} of t-\ch{PtBi2} all exhibit superconductivity under low temperature.

We begin our investigation by evaluating the feasibility of exfoliating a monolayer from the bulk  t-\ch{PtBi2}.
After fully relaxing the crystal structure, the equilibrium lattice constants of bulk t-\ch{PtBi2} are optimized to $a = b = 6.589 \ \text{\AA}$ and $c = 6.077 \ \text{\AA}$, which agree well with the experimentally measured values  \cite{PhysRevB.94.165119}.
As shown in Fig.~\ref{fig:fig1}(b), the exfoliation energy is estimated by simulating the gradual separation of the surface monolayer in a five-layer slab. 
For \ch{PtBi2}, the calculated exfoliation energy is 1.02 $\rm{J/m^2}$, which is higher than that of graphite (0.32 $\rm{J/m^2}$) \cite{exf_gra_the,exf_gra_exp} but lower than that of \ch{Ca2N} (1.31 $\rm{J/m^2}$) \cite{Ca2N}.
Both graphene and monolayer \ch{Ca2N} have been successfully isolated in experiments, suggesting that \ch{PtBi2} monolayers could also be obtained through the exfoliation process from their layered bulk crystals.

The monolayer \ch{PtBi2}, composed of a distorted Bi-Pt-Bi sandwich structure [Fig.~\ref{fig:fig1}(c)], possesses the same space group as its layered bulk phase.
The top Bi atoms are located in the same plane, while two-thirds of the Bi atoms in the bottom layer exhibit a slight vertical offset, breaking both the inversion symmetry $\mathcal{P}$ and mirror symmetry $\mathcal{M}_z$, giving rise to a perpendicular polarization $P_z$ [Fig.~\ref{fig:fig1}(c)].
For 2D systems, the out-of-plane direction is non-periodic; therefore, the polarization $P_z$ is well defined by the classical formula in electrodynamics
\begin{align}
	P_z = \frac{1}{S}\int z (\rho_{\text{ions}}+\rho_{\text{valence}}) d^3 \bm r
\end{align}
where $S$ represents the area of the unit cell. $\rho_{\text{ions}}$ and $\rho_{\text{valence}}$ denotes the charge density of ions and valence electrons, respectively.
The calculated polarization $P_z$ is 0.98 $\rm pC/m$, already higher than the polarizations of most sliding FE bilayers validated experimentally (Table \ref{tbl:polarization}) like FE metal bilayer $1\text{T}^\prime$-\ch{WTe2} \cite{yang2018origin}.

\global\long\def\arraystretch{1.4}%
\begin{table*}
	\caption{Out-of-plane polarization of common 2D ferroelectrics. Here, ML, BL and TL represent monolayer, bilayer and tetralayer, respectively.}
	\label{tbl:polarization}
	\begin{ruledtabular}
	\begin{tabular}{cccccc}
		&\ch{PtBi2} (ML) &$1\text{T}^\prime$-\ch{WTe2} (BL)& BN (BL)& \ch{MoS2}  (BL)&Graphene (TL)  \\
		\midrule
		$P_z$ ($\rm pC/m$) & 0.98 & ~0.35 \cite{yang2018origin}&2.08 \cite{li2017binary}&~0.61 \cite{MoS2_sf} & ~0.21 \cite{PhysRevLett.131.096801}  \\
	\end{tabular}
	\end{ruledtabular}
\end{table*}

\begin{figure}[htb!]
	\begin{centering}
		\includegraphics[width=\linewidth]{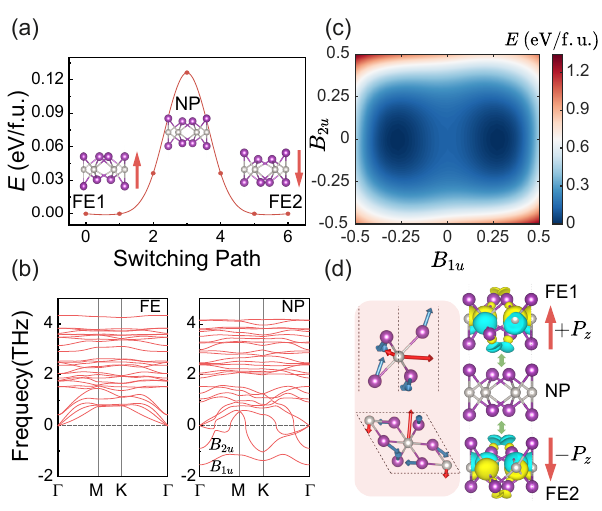}
		\par\end{centering}
	\caption{(a) Ferroelectric switching pathway of \ch{PtBi2}. (b) Phonon dispersion of FE state and NP state, where the soft modes in NP state are labeled by $B_{1u}$ and $B_{2u}$. (c) Energy landscapes as a function of ($B_{1u}$, $B_{2u}$) by freezing the modes with step-changed amplitude. (d) Top and side view of the NP structures together with the eigen-mode $B_{1u}$  marked by red and blue arrows in the left shaded area. An isosurface of the difference in charge densities of FE states (FE1 and FE2) with NP state are shown in the right panel, where light blue and yellow color denotes electron depletion and accumulation, respectively.}
	\label{fig:fig2} 
\end{figure}

As shown in Fig.~{\ref{fig:fig2}}(a), the monolayer \ch{PtBi2} possesses two FE ground states that are energetically degenerate with opposite polarizations. 
We calculate the FE switching pathway and identify a centrosymmetric non-polar (NP) state. The switching barrier of 0.126 $\rm{eV/f.u.}$ is moderate compared with $d1\rm T$-\ch{MoS2} (0.23 $\rm{eV/f.u.}$) \cite{d1TMoS2} and \ch{CuCrS2} (0.23 $\rm{eV/f.u.}$) \cite{zttCuCrS2} that have been experimentally confirmed to be ferroelectric \cite{d1TMoS2_exp,zhai_CuCrS2}, suggesting it to be a promising candidate of intrinsic 2D FE metal.    

The phonon spectra of FE and NP states are shown in Fig.~\ref{fig:fig2}(b).
In contrast to the dynamically stable FE state, the NP state exhibits two imaginary optical phonon bands.
These two modes at the $\Gamma$ point are labeled $B_{1u}$ and $B_{2u}$ which are both antisymmetric under spatial inversion operation $\mathcal P$.
These two zone-center soft modes are closely associated with FE transition \cite{PhysRevLett.3.412,Cochran,Blinc}, which is also confirmed by the 2D energy landscape as a function of ($B_{1u}$, $B_{2u}$) by freezing the modes with step-changed amplitude in Fig.~\ref{fig:fig2}(c), where $B_{1u}$ is the dominant phonon mode responsible for driving the structural transition from the NP state to FE state.
The eigen-mode of $B_{1u}$ is marked by red and blue arrow in Fig.~\ref{fig:fig2}(d) for Pt and Bi atoms, respectively.
By following the soft mode $B_{1u}$, we can obtain the two FE ground state [FE1 and FE2 in Fig.~\ref{fig:fig2}(d)] which are related by a spatial inversion operation $\mathcal P$.
The differential charge density of the FE1 and FE2 states [Fig.~\ref{fig:fig2}(d)], relative to the NP state, clearly illustrates the opposite directions of their out-of-plane polarization.

\begin{figure}[htb!]
	\begin{centering}
		\includegraphics[width=\linewidth]{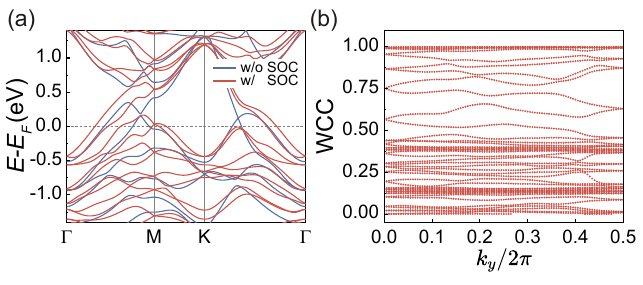}
		\par\end{centering}
	\caption{
	(a) Band structure of monolayer \ch{PtBi2} without and with SOC, depicted in blue and red colors, respectively. 
	(b) Evolution of WCCs for \ch{PtBi2} monolayer in the $k_z = 0$ plane, which indicates its topological invariant $\mathbb{Z}_2 = 1$.
	}
	\label{fig:fig3} 
\end{figure}

The electronic band structure of monolayer \ch{PtBi2}  is shown in Fig.~\ref{fig:fig3}(a), with several bands crossing the Fermi level, indicating its strong metallicity. 
Compared to $1\text{T}^\prime$-\ch{WTe2}, which is often classified as a semimetal, the \ch{PtBi2} monolayer exhibits stronger metallicity and larger polarization (Table~\ref{tbl:polarization}), attributable to its intrinsic noncentrosymmetric monolayer structure.
The SOC significantly lifting the degenerate points located along the $\Gamma$-K path induces a global gap.
Thus, the \ch{PtBi2} monolayer possesses time-reversal symmetry and maintains a continuous finite energy gap between electron-like and hole-like bands, which can be classified by the $\mathbb{Z}_2$ topological invariant \cite{KaneZ2_1, KaneZ2_2}.
Our Wannier charge center (WCC) calculations \cite{WU2018405,PhysRevB.83.235401,PhysRevB.84.075119} for the $k_z = 0$ plane  show that $\mathbb Z_2 = 1$ [Fig.~\ref{fig:fig3}(b)], revealing a topologically nontrivial electronic structure.
This nontrivial band topology may significantly enhance the interband correlation \cite{PhysRevLett.119.266804} near the Fermi level, which can manifest in interband optical transitions \cite{lijuJPCL,PhysRevLett.119.266804} and result in a giant bulk photovoltaic effect \cite{PhysRevLett.116.237402,PhysRevLett.130.256902}.

\textcolor{blue}{\textit{Ferroelectric bulk photovoltaic effect.}}---
In noncentrosymmetric crystals subjected to uniform illumination by linearly polarized light (LPL), the shift current can be expressed as 
\begin{align}\label{NLO_response}
	j^c &= 2\sigma_{ab}^c(0;\omega,-\omega)E^a(\omega)E^b(-\omega) 
\end{align}	
where $a, b, c$ are Cartesian indices and $E(\omega)$ represents the Fourier component of the electric field at angular frequency $\omega$.

In our study of the 2D topological FE metal $\mathrm{PtBi_2}$ monolayer, we focus on the NLO responses to incident LPL perpendicular to the 2D sheet. Given the difficulty in detecting out-of-plane photocurrent in such system, we restrict our analysis to $a, b, c \in \{x, y \}$ in Eq. (\ref{NLO_response}). 
Moreover, the crystal symmetry of monolayer $\mathrm{PtBi_2}$ imposes additional constraints on the photoconductivity $\sigma_{ab}^c$ which is a rank-3 tensor. 
Specifically, the mirror operation $\mathcal M_y$ leads to the vanishing of tensor components $\sigma_{xx}^y$, $\sigma_{yy}^y$ and $\sigma_{xy}^x$. 
The $\mathcal C_{3z}$ rotation further enforces the relations $\sigma_{xx}^x = -\sigma_{yy}^x = -\sigma_{xy}^y$, so we focus our calculations on the only independent component $\sigma_{xx}^x$.

When $a = b$, the photoconductivity of the shift current can be expressed in the following compact form \cite{Sipe2000,WangChong} 
\begin{align}\label{shift_current}
	&\sigma_{aa}^c(0;\omega,-\omega) =  \notag\\
	&-\frac{\pi e^3}{\hbar^2} \int\frac{d^3 k}{(2\pi)^3}\sum_{n,m}f_{nm} |r_{nm}^a|^2 R^{c;a}_{nm} \delta(\omega_{mn} - \omega) 
\end{align}
where $f_n$ is the Fermi-Dirac distribution, with $f_{nm} = f_n - f_m$ and $\hbar\omega_{mn} = \hbar\omega_{m} - \hbar\omega_{n}$ representing the occupation number and energy differences between bands $m$ and $n$, respectively.
$r_{nm}^a$ is the interband dipole matrix and the product $|r_{nm}^a|^2\delta(\omega_{mn} - \omega)$ can be interpreted as the transition rate from band $m$ to band $n$ according to the Fermi's golden rule \cite{WangChong}. 
The shift vector $R^{c;a}_{nm}$ represents the position change of a wave packet during its pumping from band $m$ to band $n$. 
It is defined as $R^{c;a}_{nm} = \frac{\partial \phi^a_{nm}}{\partial k_c} - \xi_{nn}^c + \xi_{mm}^c$, where $\phi^a_{nm}$ is the phase of $r_{nm}^a = |r_{nm}^a| e^{i\phi^a_{nm}}$, and $\bm\xi_{mm} = i \langle {u_m} |\nabla_{\bm{k}}|{u_m}\rangle$ denotes the Berry connection.

\begin{figure}
	\begin{centering}
		\includegraphics[width=0.5\textwidth]{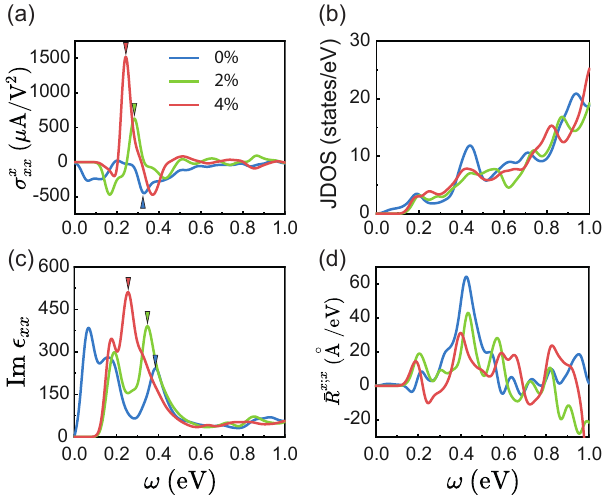}
	\par\end{centering}
	\caption{
		(a) Shift-current conductivity $\sigma_{xx}^x$, (b) JDOS, (c) the absorptive part of the dielectric function $\Im\epsilon_{xx}$ and (d) aggregate shift vector $\bar{R}^{x;x}$ as a function of the photon energy under different biaxial tensile strain.  
		\label{Fig_shiftcurrent}}
\end{figure}

Fig. \ref{Fig_shiftcurrent}(a) shows the shift current spectrum of monolayer \ch{PtBi2}, revealing a maximum over 400 $\mu\rm A /V^2$, larger than typically values in most ferroelectric materials.

Furthermore, strain engineering has been widely employed in the experimental study of 2D materials to modulate their properties \cite{InfoMat,frisenda2017biaxial,dong2023giant}.
 
Upon applying a moderate biaxial tensile strain of up to 4\%, two notable changes are observed: (i) there is a decrease in the low-frequency part of $\sigma_{xx}^x$ as the strain increases. (ii) the peak near 0.3 eV exhibits a significant enhancement accompanied by a red shift due to the tensile strain. 
To understand these two features, we calculated three additional Brillouin zone(BZ)-integrated quantities: the joint density of states (JDOS) $D_{\text{joint}}(\omega)$, the absorptive part of the dielectric function $\epsilon_{ab}(\omega)$, and the aggregate shift vector $\bar{R}^{a;b}(\omega)$ \cite{RappePRL} (see Supplemental Material\cite{suppmat} for detailed definitions).

As shown in Fig. \ref{Fig_shiftcurrent}(a) and (c), the shift current inherits most of its features from the absorption spectrum $\Im \epsilon_{xx}(\omega)$. In other words, the interband dipole matrix $r_{nm}^x$ plays a crucial role in elucidating the origins of (i)-(ii), which is further supported by the detailed calculations in Supplemental Material \cite{suppmat}.

\begin{figure}
	\begin{centering}
		\includegraphics[width=\linewidth]{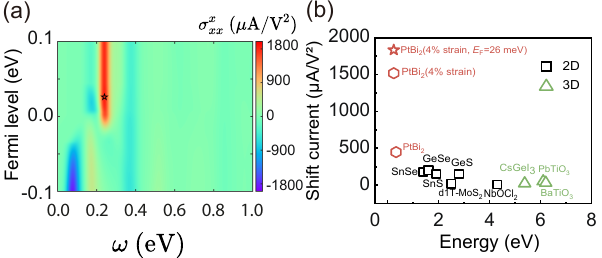}
		\par\end{centering}
	\caption{
		 (a) Shift-current conductivity $\sigma_{xx}^x$ as a function of Fermi level and photon energy under 4\% biaxial strain, with the pentagram highlighting the maximum value. (b) A comparison of maximum shift current and corresponding photon energy in \ch{PtBi2} with other widely studied 2D/3D ferroelectric materials \cite{MX_sc,d1t_mos2_sc,NbOCl2_sc,CsGeI3_sc,RappePRL}.
		}
		\label{Fig_compare}
\end{figure}

The position of Fermi level also significantly impacts the shift-current spectrum and is particularly susceptible during the synthesis of real materials. 
In Fig. \ref{Fig_compare} (a), we present the complete shift-current spectrum for a Fermi level range of -0.1 to 0.1 eV under 4\% biaxial strain (results for 0\% and 2\% strain are available in the Supplemental Material \cite{suppmat}).    
The pentagram marks the maximum of $\sigma_{xx}^x$ exceeding 1800 $\mathrm{\mu A/V^2}$, corresponding to a Fermi level of 26 meV and a photon energy of 0.24 eV, much higher compared with other widely studied 2D/3D FE materials depicted in Fig. \ref{Fig_compare}(b).

As discussed earlier, these two FE states of monolayer \ch{PtBi2} are related by an inversion operator $\mathcal P$. 
In other words, when the FE1 state ($+P_z$) is switched to FE2 state ($-P_z$) by an external electric field,  this is equivalent to applying an operation $\mathcal P$ to the FE1 state, \ie $\text{FE1} (+P_z) \xrightarrow{\mathcal P} \text{FE2} (-P_z)$. On the other hand, the shift current is a second-order nonlinear optical response expressed by Eq. (\ref{NLO_response}) where the response tensor $\sigma^{c}_{ab}$ is odd under the operation $\mathcal P$, \ie $\sigma^{c}_{ab} \xrightarrow{\mathcal P} -\sigma^{c}_{ab}$.
Therefore, it indicates that the direction of this giant shift current and FE polarization can be simultaneously reversed by an electric field \cite{kim2019prediction,xiao2022non}, making this NLO response highly tunable.

In conclusions, we theoretically predict the coexistence of electric polarization and metallicity in monolayer \ch{PtBi2}. Moreover, its topologically nontrivial electronic structure bestows upon \ch{PtBi2} with a giant ferroelectric bulk photovoltaic effect, which can be further enhanced by applying a strain, surpassing the performance of previously reported ferroelectric materials. Our findings not only provide a rare and practical material platform for the experimental study of ferroelectric metals but also highlight the tunability of the shift current  in \ch{PtBi2}, offering vast opportunities for the next generation of electronic devices.

\acknowledgments
This work was supported by the National Key R\&D Program of China (Grant No. 2020YFA0308800), the NSF of China (Grants Nos.  12234003, 12321004 and 12004035) and the China Postdoctoral Science Foundation (Grants Nos. 2021TQ0043 and 2021M700437).

\bibliography{total.bib}

\end{document}


\title
	{Supplemental Material for “Two-dimensional topological ferroelectric metal with giant shift current”}

	\author{Liu Yang}
	\affiliation
	{Department of Physics, Hubei Engineering Research Center of Weak Magnetic-field Detection, China Three Gorges University, Yichang, 443002, China}

	\author{Lei Li}
	\email[Liu Yang and Lei Li contributed equally to this work\\]{lilei1993@bit.edu.cn}
	\affiliation{Centre for Quantum Physics, Key Laboratory of Advanced Optoelectronic Quantum Architecture and Measurement (MOE), School of Physics, Beijing Institute of Technology, Beijing, 100081, China}
	\affiliation{Beijing Key Lab of Nanophotonics \& Ultrafine Optoelectronic Systems, School of Physics, Beijing Institute of Technology, Beijing, 100081, China}

	\author{Zhi-Ming Yu}
	\affiliation{Centre for Quantum Physics, Key Laboratory of Advanced Optoelectronic Quantum Architecture and Measurement (MOE), School of Physics, Beijing Institute of Technology, Beijing, 100081, China}
	\affiliation{Beijing Key Lab of Nanophotonics \& Ultrafine Optoelectronic Systems, School of Physics, Beijing Institute of Technology, Beijing, 100081, China}
	\affiliation{International Center for Quantum Materials, Beijing Institute of Technology, Zhuhai, 519000, China}

	\author{Menghao Wu}
	\email{wmh1987@hust.edu.cn}
	\affiliation
	{School of Physics, Huazhong University of Science and Technology, Wuhan, 430074, China}

	\author{Yugui Yao}
	\email{ygyao@bit.edu.cn}
	\affiliation{Centre for Quantum Physics, Key Laboratory of Advanced Optoelectronic Quantum Architecture and Measurement (MOE), School of Physics, Beijing Institute of Technology, Beijing, 100081, China}
	\affiliation{Beijing Key Lab of Nanophotonics \& Ultrafine Optoelectronic Systems, School of Physics, Beijing Institute of Technology, Beijing, 100081, China}
	\affiliation{International Center for Quantum Materials, Beijing Institute of Technology, Zhuhai, 519000, China}

	\maketitle
	
	\section{Computation methods}
	\subsection{First-principles calculations}
	Density functional theory (DFT) calculations involved in this work were implemented in the Vienna ab initio Simulation Package (VASP 5.4) code \cite{vasp1,vasp2}. 
	The exchange and correlation interactions were described by Perdew-Burke-Ernzerhof (PBE) \cite{GGA} functional based on the generalized gradient approximation (GGA).
	Projector-augmented wave (PAW) method  \cite{PAW} was adopted and the kinetic energy cutoff is set to be 400 eV. The first BZ was sampled by Monkhorst-Pack meshes method \cite{MPsampling} with a 9$\times$9$\times$1 $k$-point grid for primitive cell. The DFT-D3 functional \cite{D3} of Grimme with BJ-damping \cite{bj_damping} was used to describe the van der Waals interactions.
	A large vacuum region with a thickness of 25 $\text\AA$ was added in the $z$ direction to avoid interaction between adjacent slabs. 
	Due to the strong metallicity of monolayer \ch{PtBi2}, a second order Methfessel-Paxton smearing method method was used for structural relaxations and total energy calculation. The force and energy convergence criterion was set to  $10^{-3}~\text{eV/\AA}$ and $10^{-6} \ \rm{eV}$, respectively. 
	The dipole correction \cite{dipole_correction1,dipole_correction2} was considered during the calculation and the FE switching pathway was obtained by using the solid-state nudged elastic band (SSNEB) method \cite{NEB1,NEB2}.  
	The phonon spectrum calculations were implemented in the PHONOPY code \cite{phonopy-phono3py-JPCM,phonopy-phono3py-JPSJ} based on the finite displacement method.

	To calculate the nonlinear photoconductivity, a Wannier tight binding Hamiltonian consisting of Pt-$5d$ and Bi-$6p$ orbitals is constructed using the Wannier90 package \cite{wannier90_2020}. 
	The BZ integration is sampled with a $k$-mesh $1000\times1000\times1$ and the convergence is well tested with a denser $k$-mesh $1500\times1500\times1$.
	To obtain a 3D-like photoconductivity and eliminate the influence of vacuum space, an effective thickness of $L = 4.06 \ \text{\AA}$ is utilized.

	\subsection{Nonlinear photoconductivity}
	In the numerical calculations, the the interband dipole matrix $r_{nm}^a$ can be obtained with the relation
	\begin{align}
		r_{nm}^a = \frac{v_{nm}^a}{i\omega_{nm}} (n \neq m)  
	\end{align}
	where $v_{nm}^a   = \frac{1}{\hbar}\bra{n}\frac{\partial H}{\partial k_a}\ket{m}$ is interband velocity matrix element. In Eq. (3) in the main text, we only need the real part of the shift vector $R^{c;a}_{nm}$ \cite{Sipe2000,PhysRevXNagaosa}, which can be calculated as 
	\begin{align}
		\mathrm{Re}\left[ R^{c;a}_{nm} \right] = \mathrm{Im} \frac{1}{v_{nm}^a}\left[ \frac{v_{nm}^a\Delta_{nm}^c + v_{nm}^c\Delta_{nm}^a}{\omega_{nm}} - W_{nm}^{ac} + \sum_{p \neq n,m}\left(\frac{v_{np}^a v_{pm}^c}{\omega_{pm}} - \frac{v_{np}^c v_{pm}^a}{\omega_{np}} \right) \right](n\neq m)  
	\end{align}
	where $W_{nm}^{ac} = \frac{1}{\hbar}\bra{n}\frac{\partial^2 H}{\partial k_a\partial k_c}\ket{m}$ and $\Delta_{nm}^a = v_{nn}^a - v_{mm}^a$ is the velocity difference between bands $n$ and $m$. 
	After obtaining $v_{nm}^a$, $r_{nm}^a$, and $R^{c;a}_{nm}$ from the Wannier interpolations, we can calculate the nonlinear photoconductivity in Eq. (3) and other BZ-integrated quantities. The JDOS $D_{\text {joint }}(\omega)$, dielectric function $\epsilon_{a b}(\omega)$ and aggregate shift vector $\bar{R}^{a;b}(\omega)$ \cite{RappePRL} in the main text are defined as 
	\begin{align}
		D_{\text {joint }}(\omega) &= \frac{\Omega}{\hbar} \int\frac{d^3 k}{(2\pi)^3} \sum_{n, m} f_{n m} \delta\left(\omega_{m n}-\omega\right) \\
		\epsilon_{a b}(\omega) &= \frac{i \pi e^2}{\epsilon_0 \hbar} \int\frac{d^3 k}{(2\pi)^3} \sum_{n, m} f_{n m} r_{n m}^a r_{m n}^b \delta\left(\omega_{m n}-\omega\right) \\
		\bar{R}^{a;b}(\omega) &= \frac{\Omega}{\hbar} \int\frac{d^3 k}{(2\pi)^3} \sum_{n, m} f_{n m} R^{a;b}_{n m}\delta\left(\omega_{m n}-\omega\right)
	\end{align}	
	where $\Omega$ is the volume of the unit cell.

	\section{The origin of strain-enhanced shift current}

	\begin{figure}
		\begin{centering}
			\includegraphics[width=\textwidth]{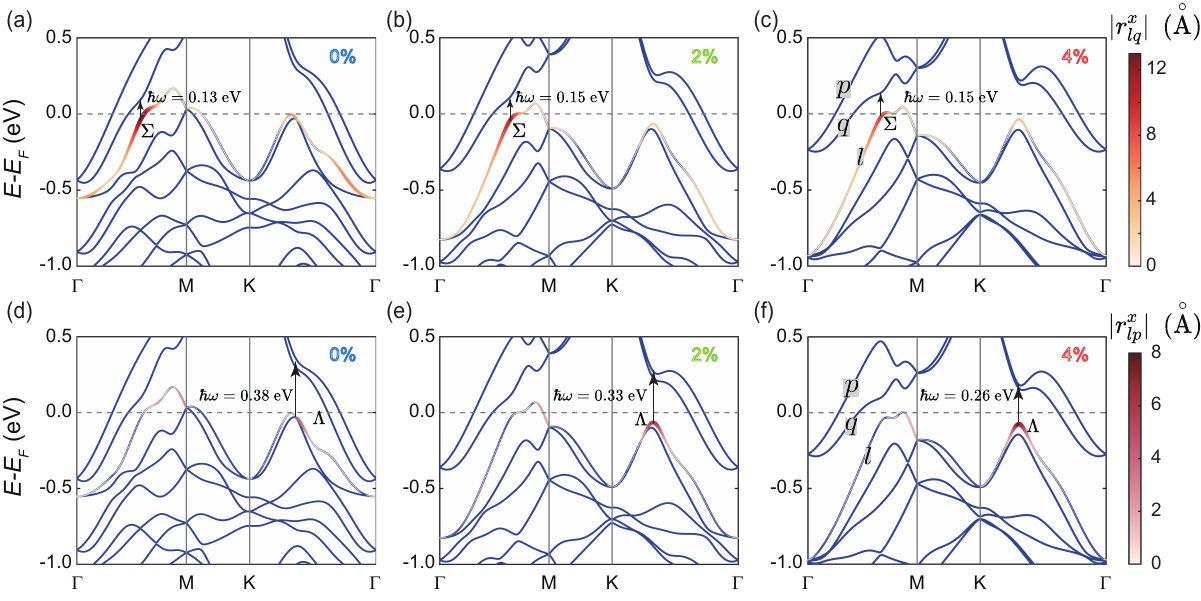}
			\par\end{centering}
		\caption{
			Band structures under 0\%, 2\%, and 4\% biaxial strain, with (a)-(c) mapped with $k$-resolved $|r_{lq}^x|$ and (d)-(f) mapped with $k$-resolved $|r_{lp}^x|$.
			\label{Fig_strain}}
	\end{figure}

	Band transitions near the Fermi level provide the major contribution to the shift current and absorption spectra, so here we confine our analysis to three bands, labeled as $p$, $q$, and $l$, as illustrated in Fig. \ref{Fig_strain} (c) and (f).
	The $k$-resolved $|r_{lq}^x|$ and $|r_{lp}^x|$ with different strain are mapped onto the corresponding band structures in Fig. \ref{Fig_strain}.  
	It can be observed that the distribution of $|r_{lq}^x|$ is primarily concentrated near the $\Sigma$ point, with its magnitude decreasing as the strain increases. 
	Considering the energy difference between band $l$ and $q$ at $\Sigma$, the observed behavior of $|r_{lq}^x|$ can explain the reduction in $\sigma_{xx}^x$ at low frequency.  
	By contrast, the distribution of $|r_{lp}^x|$ is mainly localized around the $\Lambda$ point, where its magnitude increases with strain. 
	On the other hand, the energy difference at $\Lambda$  is reduced from 0.38 eV to 0.26 eV as a result of the strain-induced transformation in the band structure's shape.
	Therefore, the changes (i) and (ii)  in shift-current spectrum can be reasonably attributed to the strain-induced variation of $|r_{lq}^x|$ and $|r_{lp}^x|$, respectively.

	\section{Fermi Level Dependence of the Shift-Current Spectrum}

\begin{figure}
		\begin{centering}
			\includegraphics[width=\textwidth]{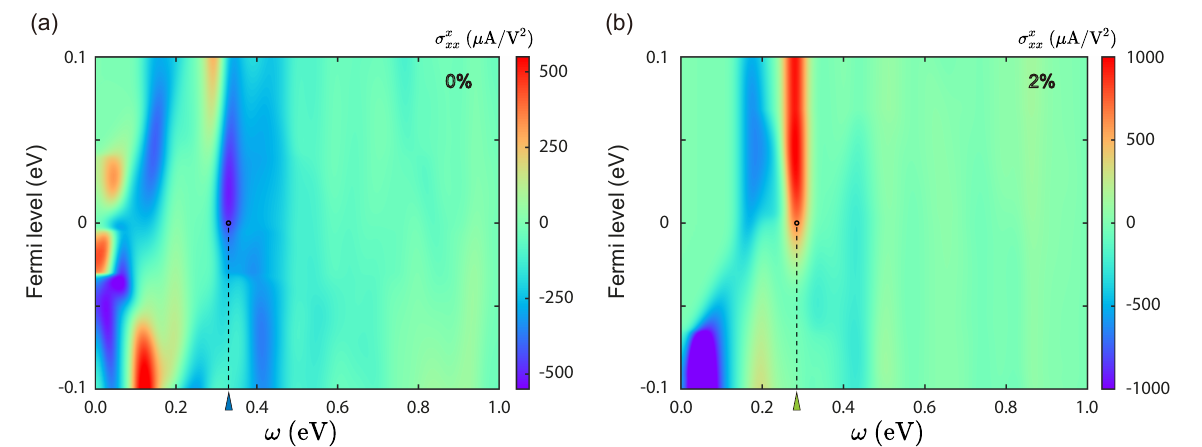}
			\par\end{centering}
		\caption{
			Shift-current spectrum under (a) 0\% and (b) 2\% biaxial strain as a function of Fermi level and photon energy.
			\label{Fig_Fermi}}
	\end{figure}

	To evaluate the impact of the Fermi level on the shift-current spectrum under 0\% and 2\%  biaxial strain, we calculated the complete spectrum for Fermi levels range from -0.1 to 0.1 eV, with the results presented in Fig. \ref{Fig_Fermi}.
	For the unstressed \ch{PtBi2} monolayer, the low-frequency part of the spectrum is enhanced by slight hole doping, while the peak near 0.3 eV is amplified upon slight electron doping.
	Under a 2\% strain, the peak near 0.3 eV increase to more than 1000 $\mathrm{\mu A/V^2}$  when electron-doped.
	Overall, our results demonstrate that modifying the Fermi level is an effective tool for tuning the photovoltaic response in the \ch{PtBi2} monolayer, with enhanced photovoltaic characteristics observed under specific doping conditions.

	\bibliography{total.bib}